\documentclass[prc,preprint,floatfix,nofootinbib,tightenlines,%
               superscriptaddress]{revtex4}
\usepackage{amsmath,amssymb}

\newcommand{\beqn}{\begin{equation}}
\newcommand{\eeqn}{\end{equation}}
\newcommand{\bea}{\begin{eqnarray}}
\newcommand{\eea}{\end{eqnarray}}
\newcommand{\ba}{\begin{align}}
\newcommand{\ea}{\end{align}}

\newcommand{\la}{\Lambda}

\newcommand{\vlowk}{V_{{\rm low}\,k}}

\newcommand{\tlowk}{T_{{\rm low}\,k}}

\newcommand{\Tmatrix}{$T$-matrix}

\begin{document}

\title{Comment on ``Problems in the derivations of the renormalization
       group equation for the low momentum nucleon interactions''}

\author{S.K.\ Bogner}
\affiliation{National Superconducting Cyclotron Laboratory and Department 
  of Physics and Astronomy, Michigan State University, East Lansing, MI 48844}
\author{R.J.\ Furnstahl}
\affiliation{Department of Physics, The Ohio State University, Columbus, OH 43210}
\author{A.\ Schwenk}
\affiliation{TRIUMF, 4004 Wesbrook Mall, Vancouver, BC, Canada, V6T 2A3}


\begin{abstract}
The criticisms in manuscript arXiv:0803.4371 do not change the 
renormalization group equation for the ``$\vlowk$'' interaction
nor do they affect any results in the literature.  Several other
potentially misleading statements about low-momentum interactions 
are also addressed.
\end{abstract}

\maketitle

A recent paper~\cite{Harada:2008am} makes various statements about
renormalization group (RG) equations for low-momentum interactions
(``$\vlowk$'') as derived by Bogner {\it et al.} in Ref.~\cite{ref14}.
The earlier posted versions of the manuscript 
(v1 and v2 of Ref.~\cite{Harada:2008am}) claimed that the RG equations
were incorrect because of a missing term.
After we communicated to the author an explicit demonstration 
that this term vanishes, as well as
a straightforward proof that the conventional model-space methods 
must obey the same RG equation, the manuscript was modified 
(v3 and v4 of Ref.~\cite{Harada:2008am}) 
to criticize only the derivations in Ref.~\cite{ref14} 
rather than the final RG equation 
or the equivalence of the model-space and RG approaches. 
While we agree that the original derivations in Ref.~\cite{ref14}
lack formal mathematical rigor, 
we feel that readers of 
Ref.~\cite{Harada:2008am} could still be 
misled that there are problems with $\vlowk$. Therefore
we have written this comment.

We emphasize in particular the following point:
\emph{None of the calculations in the literature involving $\vlowk$ 
are affected by the discussion in Ref.~\cite{Harada:2008am}.}
Low-momentum nucleon-nucleon interactions can be derived and constructed
in energy-independent form either using model-space methods
(such as Lee-Suzuki or Okubo) or through an RG treatment.
To our knowledge,
there is no question about the validity of the former; the issue
in Ref.~\cite{Harada:2008am} concerned the
precise equivalence of the approaches. However, in v3 and v4 the author
acknowledges that this equivalence is indeed correct.
In practice, most calculations using $\vlowk$ have used the model-space
methods rather than the differential equations of the RG
because they are more robust numerically. This is true for
both sharp and smooth cutoff versions of $\vlowk$. 
Thus doubts in Ref.~\cite{Harada:2008am} about the final RG 
equation have no impact on prior results.   

Nevertheless, the equivalence of the model-space and RG methods is important
conceptually.  Reference~\cite{ref14} proposed four derivations,
none of which is rigorous in a mathematical sense.  The most straightforward
issue pointed out in Ref.~\cite{Harada:2008am}, which could
undermine two of the derivations, is that 
the use of the completeness relation
did not separately treat the bound states, as is usually
required.  We reproduce our original argument here that such a treatment,
as carried out in Ref.~\cite{Harada:2008am}, does not
in the end lead to changes in the RG equation proposed in
Ref.~\cite{ref14}.
We would welcome a more rigorous mathematical treatment of this
and other aspects of
the $\vlowk$ RG equations, but leave that to others.

The low-momentum partial-wave
interaction $\vlowk$~\cite{Vlowk} is defined by the following
equation for the half-on-shell (HOS) \Tmatrix\ 
(in units where $\hbar^2/m = 1$), 
\beqn
\label{Eq:cutT}
\tlowk(k',k;k^2) = \vlowk(k',k) + \frac{2}{\pi}\int_0^\la p^2dp\, 
\frac{\vlowk(k',p)\tlowk(p,k;k^2)}{k^2 - p^2} \;,
\eeqn
where Cauchy principal value integrals are implied throughout and  the
quantum numbers labeling the partial wave have been suppressed.
Demanding $\frac{d}{d\la}\tlowk(k',k;k^2) = 0$ implies 
an RG equation,  which is claimed in
Ref.~\cite{ref14} and elsewhere to be 
\beqn
\label{Eq:OurRGE}
\frac{d}{d\la}V^\la(k',k) = 
\frac{2}{\pi}\frac{V^\la(k',\la)T^\la(\la,k;\la^2)}{1-(k/\la)^2}\;.
\eeqn
Here we have switched to the abbreviated notation $V^\la\equiv\vlowk$.  
By construction, the low-momentum HOS \Tmatrix\  of the
initial interaction is preserved by the interaction $V^\la$ obtained
from integrating the RG equation.
  
Reference~\cite{Harada:2008am} states that the 
RG equation, Eq.~(\ref{Eq:OurRGE}), must be
modified in the event that bound states are present ({\it i.e.}, in
the $^3$S$_1$--$^3$D$_1$ channel) to
\beqn
\label{Eq:HaradaRGE}
\frac{d}{d\la}V^\la(k',k) = \frac{2}{\pi}\frac{V^\la(k',\la)
  T^\la(\la,k;\la^2)}{1-(k/\la)^2} + \delta\beta^\la(k',k)\;.
\eeqn
The extra term arising from the bound states is given
by~\cite{Harada:2008am}
\begin{eqnarray}
  \label{Eq:extraterms}
    \delta\beta^{\la}(k',k) &=& \frac{2}{\pi}\sum_{i}\int_0^{\Lambda} l^2dl\, 
    \Biggl( \frac{d}{d\Lambda}V^\la(k',l) -\frac{2}{\pi}
    \frac{V^\la(k',\Lambda)V^\la(\Lambda,l)}{1 + (k_{B_i}/\Lambda)^2}\Biggr) 
    \chi^\la_{B_i}(l)\bigl(\widetilde{\chi}^\la_{B_i}(k)\bigr)^* \;, \nonumber \\
  &=& \sum_i\Biggl(\langle k'|\frac{dV^{\la}}{d\Lambda}|\chi^\la_{B_i}\rangle
  -\frac{2}{\pi}\frac{\langle
    k'|V^\la|\Lambda\rangle\langle\Lambda|V^\la|\chi^\la_{B_i}\rangle}{1 +
  (k_{B_i}/\Lambda)^2}\Biggr)\bigl(\widetilde{\chi}^\la_{B_i}(k)\bigr)^*\;,
\end{eqnarray}
where the bound state wave functions 
$\chi^\la_{B_i}(k) \equiv \langle k|\chi^\la_{B_i}\rangle$ satisfy 
\begin{equation}
  \bigl (H_0 + V^{\la}\bigr) |\chi^\la_{B_i}\rangle 
  = -k_{B_i}^2|\chi^{\la}_{B_i}\rangle\;,
\end{equation}
and we have used the plane wave completeness relation $P =
\frac{2}{\pi} \int_0^\la p^2dp\, |p\rangle\langle p|$ to obtain the
second line in Eq.~(\ref{Eq:extraterms}).

We now provide a simple demonstration that the $\delta\beta^\la(k',k)$ term is
identically zero. As a consequence of the cutoff
independence of the HOS \Tmatrix, 
the interacting scattering eigenstates of the
low-momentum Hamiltonian $H^\la = H_0 + V^{\la}$ are equal to the low-momentum
projections of the corresponding eigenstates of the ``bare''
Hamiltonian, $|\chi_k\rangle = P|\Psi_k\rangle$, with an
analogous relation for bound states, $|\chi^{\la}_{B_i}\rangle = P|\Psi_{B_i}\rangle$
or equivalently
$\frac{d}{d\Lambda}\langle k'|V^{\la}|\chi^{\la}_{B_i}\rangle=0$.
Moreover, the preservation of the low-momentum part of the bare HOS
\Tmatrix\  also implies that bound state poles of the input Hamiltonian
should be preserved by $H^\la$ (provided that the binding
momentum is less than the sharp cutoff). With the aid of these
observations we find
\begin{eqnarray}
 \frac{d}{d\Lambda} k_{B_i}^2 &=& 0 \;, \\
 \frac{d}{d\Lambda}|\chi^\la_{B_i}\rangle &=& 
 \frac{2}{\pi}\Lambda^2|\Lambda\rangle\langle\Lambda|\Psi_{B_i}\rangle 
  =
 \frac{2}{\pi}\Lambda^2|\Lambda\rangle\langle\Lambda|\chi^\la_{B_i}\rangle
  \;.
\end{eqnarray}
Taking $d/d\Lambda$ of both sides of the bound state Schr\"odinger
equation and making use of the previous two equations leads to
\begin{equation}
\frac{dV^\la}{d\Lambda}|\chi^\la_{B_i}\rangle 
   = -\frac{2}{\pi}\Lambda^2 
   \bigl(k_{B_i}^2 + H_0+ V^\la\bigr)
   |\Lambda\rangle\langle\Lambda|\chi^\la_{B_i}\rangle\;.
\end{equation}
Multiplying from the left by a $P$-space plane wave state $\langle k'|$
gives an expression for the first term in the brackets in
Eq.~(\ref{Eq:extraterms}),
\begin{equation}
\langle k'|\frac{dV^\la}{d\Lambda}|\chi^\la_{B_i}\rangle 
  = -\frac{2}{\pi}\Lambda^2 \langle k'
  |V^\la|\Lambda\rangle\langle\Lambda|\chi^\la_{B_i}\rangle\;.
\end{equation}
Finally, making the trivial substitution $V^\la = H^\la - H_0$ in the
second term in Eq.~(\ref{Eq:extraterms}) gives
\begin{eqnarray}
   -\frac{2}{\pi}\frac{\langle k'|V^\la|\Lambda\rangle\langle\Lambda
   |V^\la|\chi^\la_{B_i}\rangle}{1 + (k_{B_i}/\Lambda)^2} 
   &=& -\frac{2}{\pi}\frac{\langle k'|V^\la|\Lambda\rangle\langle\Lambda
     |\bigl(H^\la-H_0)|\chi^\la_{B_i}\rangle}{1 + (k_{B_i}/\Lambda)^2} \\
  &=&\frac{2}{\pi} \Lambda^2 \langle k'|V^\la|
    \Lambda\rangle\langle\Lambda|\chi^\la_{B_i}\rangle\;,
\end{eqnarray}
which evidently cancels the first term.  Therefore, we have shown that
$\delta\beta^\la(k',k) = 0$, and the original RG equation,
Eq.~(\ref{Eq:OurRGE}), holds independent of the presence of bound
states. The equivalence of the model-space methods
to the RG equation follows immediately, since it is well-known that
such methods preserve the HOS \Tmatrix, the bound state poles, and
the $P$-space projection of low energy eigenstates.   
(In addition, it follows directly that the extra terms in Eq.~(4.8)
of Ref.~\cite{Harada:2008am} are cutoff independent.)
By the same line of reasoning the smooth-cutoff version of the 
RG equation from Ref.~\cite{Bogner:2006vp} 
is also unmodified by bound states. 

Given the subtleties that can arise in any such demonstrations,
a practical test is whether actual numerical applications
of the RG equations reveal any problems related to completeness
in channels with bound states. 
This test is not ideal because 
the numerical implementation of the RG equations is difficult,
and errors for phases shifts or bound-state energies
grow in each channel as the cutoff is lowered.
However, there is no sign of special problems or enhanced errors in the
$^3$S$_1$--$^3$D$_1$ channel~\cite{KaiSuna}. 

There are several other points in the manuscript that we feel
need clarification or correction.
The first is the relationship of the RG evolution and chiral effective
field theory (EFT) interactions.
If one starts from a chiral 
EFT interaction determined to a given order in the power 
counting, the 
truncation error of the initial chiral EFT is exactly preserved when 
many-body interactions are evolved as well.
If many-body interactions are truncated, we find
that the error remains
of the same natural size for cutoffs used in practical calculations.
Therefore, 
$\vlowk$ preserves the systematic nature of the initial EFT.
On the other hand, if 
one starts from a phenomenological interaction, then 
it is correct that
the low-momentum Hamiltonian is not systematically improvable 
since the starting point is not either.

There are also comments on the smooth-cutoff version of $\vlowk$ that may
leave a wrong impression.
In particular, the three-step numerical treatment applied in
Ref.~\cite{Bogner:2006vp}, which is used in practice, is numerically robust.
While analytic analysis may be difficult, the converse is true
for energy-dependent Wilsonian RG implementations, whose simplicity is
highlighted in Ref.~\cite{Harada:2008am}.
That is, while it is much easier to perform a fixed-point analysis
within the energy-dependent framework,
it is notoriously difficult to work with in few- and 
many-body calculations (to our knowledge no application to $A>3$ exists).

Finally, there are comments in Ref.~\cite{Harada:2008am} 
on the Similarity Renormalization Group (SRG) approach (inexplicably
called WGW in Ref.~\cite{Harada:2008am}) about the non-invariance
of the HOS \Tmatrix\  under the SRG, which seem to imply this poses a
problem. We agree that the HOS \Tmatrix\  changes with SRG evolution, but 
disagree that this is a problem. The HOS \Tmatrix\  is certainly not an 
observable, while one can easily show that the on-shell \Tmatrix\  and
bound-state poles are invariant under the SRG 
evolution. Indeed, the invariance of observables 
follows trivially from the fact that the SRG
evolution represents a series of unitary transformations. 

\begin{acknowledgments}
This work was supported in part by the National Science 
Foundation under Grant Nos.~PHY--0354916 and PHY--0653312,  
and the Natural Sciences and Engineering Research Council of Canada 
(NSERC). TRIUMF receives federal funding via a contribution agreement 
through the National Research Council of Canada. 
\end{acknowledgments}


\end{document}